\documentclass[apjl]{emulateapj}

\usepackage{graphicx}
\usepackage{ifpdf}
\usepackage{ifthen}

\newboolean{emulateapj}
\setboolean{emulateapj}{true}

\newcommand{\ctbd}[1]{}

\newcommand{\lc}{light curve}

\newcommand{\kms}{\ensuremath{\rm km\,s^{-1}}}
\newcommand{\ms}{\ensuremath{\rm m\,s^{-1}}}
\newcommand{\mss}{\ensuremath{\rm m\,s^{-2}}}
\newcommand{\gcmc}{\ensuremath{\rm g\,cm^{-3}}}

\newcommand{\rhk}{\ensuremath{R^{\prime}_{HK}}}
\newcommand{\logrhk}{\ensuremath{\log\rhk}}

\newcommand{\teff}{\ensuremath{T_{\rm eff}}}

\newcommand{\vsini}{\ensuremath{v \sin{i}}}

\newcommand{\rsun}{\ensuremath{R_\sun}}
\newcommand{\msun}{\ensuremath{M_\sun}}
\newcommand{\lsun}{\ensuremath{L_\sun}}

\newcommand{\rstar}{\ensuremath{R_\star}}

\newcommand{\loggstar}{\ensuremath{\log{g_\star}}}
\newcommand{\mstar}{\ensuremath{M_\star}}
\newcommand{\lstar}{\ensuremath{L_\star}}

\newcommand{\rpl}{\ensuremath{R_{p}}}
\newcommand{\mpl}{\ensuremath{M_{p}}}

\newcommand{\gpl}{\ensuremath{g_{p}}}

\newcommand{\rjup}{\ensuremath{R_{\rm J}}}
\newcommand{\mjup}{\ensuremath{M_{\rm J}}}

\newcommand{\rjuplong}{\ensuremath{R_{\rm Jup}}}
\newcommand{\mjuplong}{\ensuremath{M_{\rm Jup}}}

\newcommand{\figr}[1]{Fig.~\ref{fig:#1}}

\newcommand{\tabr}[1]{\mbox{Table~\ref{tab:#1}}}

\newcommand{\flwof}{\mbox{FLWO 1.2 m}}

\newcommand{\flwos}{\mbox{FLWO 1.5 m}}

\defcitealias{kovacs05}{KBN05}
\defcitealias{fortney06}{FMB06}

\newcommand{\hatcur}{HAT-P-6}
\newcommand{\hatcurb}{HAT-P-6b}
\newcommand{\gschatcur}{GSC~03239-00992}
\newcommand{\twomasshatcur}{2MASS~23390581+4227575}

\newcommand{\hatcurm}{\ensuremath{1.06\pm0.12}}		
\newcommand{\hatcurmshort}{\ensuremath{1.06}}
\newcommand{\hatcurmlong}{\ensuremath{1.057\pm0.119}}	
\newcommand{\hatcurr}{\ensuremath{1.33\pm0.06}}		
\newcommand{\hatcurrshort}{\ensuremath{1.33}}		
\newcommand{\hatcurrlong}{\ensuremath{1.330\pm0.061}}	
\newcommand{\hatcurrho}{\ensuremath{0.558\pm0.047}}
\newcommand{\hatcuri}{\ensuremath{85\fdg51\pm0\fdg35}}
\newcommand{\hatcurg}{\ensuremath{14.8\pm1.0}}
\newcommand{\hatcurar}{\ensuremath{7.69\pm0.22}}
\newcommand{\hatcurarel}{\ensuremath{0.05235\pm0.00087}}
\newcommand{\hatcurP}{\ensuremath{3.852985\pm0.000005}}
\newcommand{\hatcurPprec}{\ensuremath{3.852985}}
\newcommand{\hatcurPshort}{3.8529}
\newcommand{\hatcurK}{\ensuremath{115.5\pm4.2}}
\newcommand{\hatcurT}{\ensuremath{2,\!454,\!035.67575\pm0.00028}}
\newcommand{\hatcurMT}{\ensuremath{54,\!035.67575\pm0.00028}}
\newcommand{\hatcurdur}{\ensuremath{0.1461\pm0.0017}}
\newcommand{\hatcuringdur}{\ensuremath{0.0188\pm 0.0011}}

\newcommand{\hatcurrprstar}{\ensuremath{0.09338\pm0.00053}}

\shorttitle{\hatcurb: Transiting Hot Jupiter around a bright F star}
\shortauthors{Noyes et al.}

\begin{document}

\ifthenelse{\boolean{emulateapj}}{
\title{\hatcurb:
	A Hot Jupiter transiting a bright F star
	\altaffilmark{$\dagger$}}}
{\title{\hatcurb:
	A Hot Jupiter transiting a bright F star
	\altaffilmark{\dagger}}}
\author{
	R.~W.~Noyes\altaffilmark{1},
	G.~\'A.~Bakos\altaffilmark{1,2},
	G.~Torres\altaffilmark{1},
	A.~P\'al\altaffilmark{1,3},
	G\'eza~Kov\'acs\altaffilmark{4},
	D.~W.~Latham\altaffilmark{1},
	J.~M.~Fern\'andez\altaffilmark{1},
	D.~A.~Fischer\altaffilmark{5},
	R.~P.~Butler\altaffilmark{6},
	G.~W.~Marcy\altaffilmark{7},
	B.~Sip\H{o}cz\altaffilmark{3,1},
	G.~A.~Esquerdo\altaffilmark{1},
	G\'abor~Kov\'acs\altaffilmark{1},
	D.~D.~Sasselov\altaffilmark{1},
	B.~Sato\altaffilmark{8},
	R.~Stefanik\altaffilmark{1},
	M.~Holman\altaffilmark{1},
	J.~L\'az\'ar\altaffilmark{9},
	I.~Papp\altaffilmark{9} \&
	P.~S\'ari\altaffilmark{9}
}
\altaffiltext{1}{Harvard-Smithsonian Center for Astrophysics,
Cambridge, MA, rnoyes@cfa.harvard.edu}
\altaffiltext{2}{Hubble Fellow}
\altaffiltext{3}{Department of Astronomy,
	E\"otv\"os Lor\'and University, Budapest, Hungary.}
\altaffiltext{4}{Konkoly Observatory, Budapest, 
	Hungary}
\altaffiltext{5}{Department of Physics and Astronomy, San Francisco
	State University, San Francisco, CA}
\altaffiltext{6}{Department of Terrestrial Magnetism, Carnegie
	Institute of Washington, DC}
\altaffiltext{7}{Department of Astronomy, University of California,
	Berkeley, CA}
\altaffiltext{8}{Tokyo Institute of Technology, 
	Tokyo, Japan}
\altaffiltext{9}{Hungarian Astronomical Association, Budapest, 
	Hungary}
\altaffiltext{$\dagger$}{
    Based in part on observations obtained at the W.~M.~Keck
    Observatory, which is operated by the University of California and
    the California Institute of Technology. Keck time has been in part
    granted by NASA (run N162Hr) and NOAO (run A285Hr).
}

\begin{abstract}
	In the ongoing HATNet survey we have detected a giant
	planet, with radius $\hatcurr\,\rjuplong$ and mass
	$\hatcurm\,\mjuplong$, transiting the bright ($V=10.5$) star
	\gschatcur{}. The planet is in a circular orbit with period
	$\hatcurP$\,days and mid-transit epoch \hatcurT\,(HJD). The parent
	star is a late F star with mass $1.29 \pm 0.06\,\msun$, radius
	$1.46 \pm 0.06\,\rsun$, $\teff \sim 6570\pm80\,\mathrm{K}$,
	$\mathrm{[Fe/H]} = -0.13 \pm 0.08$ and age $\sim
	2.3^{+0.5}_{-0.7}\,\mathrm{Gy}$. With this radius and mass,
	\hatcurb{} has somewhat larger radius than theoretically expected. 
	We describe the observations and their analysis to determine
	physical properties of the \hatcur{} system, and briefly discuss
	some implications of this finding.
\end{abstract}


\section{Introduction}
\label{sec:introduction}
The detection of transiting exoplanets is very important to exoplanet
research because of the information about both planetary radius and
mass that comes from photometric transit light curves combined with
follow-up radial velocity observations.  The transiting exoplanets
known as of this writing span a wide range in the physical parameter
space of planetary mass, radius, orbital period, semi-major axis,
eccentricity; and parent star parameters, including mass, radius,
effective temperature, metallicity, and age. Filling out their
distribution in this multidimensional space is certain to give us
important information on the origin and evolution of exoplanetary
systems. Here we report on the discovery by HATNet of its sixth
transiting planet, \hatcurb{}, an inflated Jupiter-mass gas giant in an
essentially circular orbit about an F dwarf star with slightly
sub-solar metallicity.

\section{Photometric detection}
\label{sec:detection}

The HATNet telescopes \mbox{HAT-6} and \mbox{HAT-9}
\citep[HATNet;][]{Bakos:02, Bakos:04} observed HATNet field G161,
centered at $\alpha = 00^{\rm h} 32^{\rm m}$, $\delta = +37\arcdeg
30\arcmin$, on a near-nightly basis from 2005 August 12 to 2005
December 16. We have gathered altogether 9550 5-min exposures, each
yielding photometric measurements for approximately $35,\!000$ stars
with $I<14$ and about $10,\!000$ stars with better than 2\% \lc\ rms.
The field was observed in network mode, whereby at the end of its
nightly observing sequence the \mbox{HAT-6} telescope in Arizona handed
off to the \mbox{HAT-9} telescope in Hawaii, thus extending the
duration of continuous observations. Following standard frame
calibration procedures, astrometry was performed as described in
\citet{Pal:06}, and aperture photometry results were subjected to
External Parameter Decorrelation \citep[EPD, described briefly
in][]{Bakos:07a}, and the Trend Filtering Algorithm
\citep[TFA;][]{Kovacs:05}. We searched the light curves for box-shaped
transit signals using the BLS algorithm of \citet{Kovacs:02}. A very
significant periodic dip in intensity was detected in the $I\approx
10.6$ magnitude star \gschatcur{} (also known as \twomasshatcur{},
with a depth of $9$ mmag, a period of $P=\hatcurPshort\,\mathrm{days}$,
and a duration of 3.1 hours.

\begin{figure}[!ht]
\ifpdf
\plotone{img/f1.pdf}
\else
\plotone{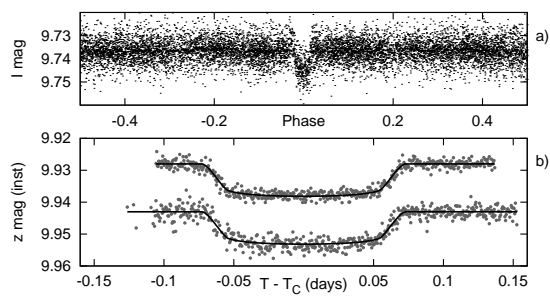}
\fi
\caption{
	(a) Unbinned instrumental $I$-band discovery light curve of
	\hatcur{} obtained with HATNet, folded with the period of $P =
	\hatcurPprec$ days. Superimposed (larger dots) is the same data
	binned to 1/200 in phase. (b) Unbinned instrumental Sloan $z$-band
	photometry collected with the KeplerCam at the \flwof\ telescope
	on 2006 October 26 (top curve) and again on UT 2007 September 4
	(next curve); superimposed on both is our best-fit transit model
	curve (see text).
\label{fig:lc}}
\end{figure}

\notetoeditor{This is the intended place of \figr{lc}. We would like to
typeset it as a single column figure.}

The HATNet discovery light curve is shown in Fig.~\ref{fig:lc}a.  As is
shown in the following sections we deduce that the signal is due to the
transit of a Jovian planet across the face of the star. Hereafter we
refer to the star as \hatcur{}, and to the planetary companion as
\hatcurb{}.

\section{Follow-up observations}
\label{sec:followup}

\hatcur{} was observed spectroscopically with the CfA Digital
Speedometer \citep{Latham:92} at the \flwos\ Tillinghast reflector
of the Fred L.~Whipple Observatory (FLWO) in order to rule out the
possibility that the observed drop in brightness is caused by a
transiting low-mass stellar companion rather than a planet, as well as
to characterize the rotation and surface gravity of the star. Seven
spectra were obtained over an interval of 92 days.  Radial velocities
were obtained by cross-correlation and have a typical precision of
$0.4\,\kms$. They showed no variation within the uncertainties, ruling
out a companion of stellar mass. The mean heliocentric radial velocity
is $-22.7 \pm 0.5$\,\kms.

Photometric follow-up of \hatcur{} was then carried out in the Sloan
$z$-band with KeplerCam \citep[see e.g.][]{Holman:07} on the \flwof\
telescope, on 2006 October 26.  An astrometric solution between the
individual frames and the 2MASS catalog was carried out using first
order polynomials based on $\sim$400 stars per frame. Aperture
photometry was performed using a series of apertures in fixed positions
around the 2MASS-based $(x,y)$ pixel coordinates. We selected a frame
taken near the meridian and used $\sim$260 stars and their magnitudes
as measured on this reference frame to transform all other frames to a
common instrumental magnitude system. The aperture yielding the lowest
scatter outside of transit was used in the subsequent analysis. The
light curve was then de-correlated against trends using the
out-of-transit sections and a dependence on hour angle (see
Fig.~\ref{fig:lc}b, upper curve).

A follow-up KeplerCam observation was recently obtained (2007 Sep 4),
for the purpose of improving the photometric accuracy of the transit
curve model, and also to determine the orbital period and mid-transit
time with maximum accuracy.  The data were treated identically to those
for the 2006 October 26 transit; results are shown in the lower curve
of Fig.~\ref{fig:lc}b.

\begin{deluxetable}{lrrc}
\tablewidth{0pc}
\tablecaption{Relative radial velocity measurements of \hatcur{}\label{tab:rvs}}
\tablehead{
	\colhead{BJD} &
	\colhead{RV} &
	\colhead{\ensuremath{\sigma_{\rm RV}}}\\
	\colhead{\hbox{~~~~(2,400,000$+$)~~~~}} &
	\colhead{(\ms)} &
	\colhead{(\ms)}
}
\startdata
54022.70228\dotfill &  $+     74.01$ &   $4.85$ & \\
54023.78901\dotfill &  $+     47.70$ &   $4.85$ & \\
54085.81512\dotfill &  $-      3.70$ &   $5.72$ & \\
54130.72257\dotfill &  $+     62.40$ &   $6.93$ & \\
54247.11201\dotfill &  $+     67.25$ &   $4.55$ & \\
54248.08955\dotfill &  $-    110.02$ &   $4.91$ & \\
54249.08558\dotfill &  $-     94.78$ &   $4.23$ & \\
54250.12528\dotfill &  $+     66.22$ &   $4.61$ & \\
54251.08766\dotfill &  $+     56.10$ &   $4.21$ & \\
54258.11601\dotfill &  $+    103.71$ &   $4.09$ & \\
54279.09485\dotfill &  $-    127.75$ &   $6.05$ & \\
54286.12699\dotfill &  $-     18.23$ &   $6.84$ & \\
54319.11890\dotfill &  $+     32.41$ &   $4.30$ & \\
54336.87654\dotfill &  $-    112.33$ &   $4.13$ & \\
54337.86117\dotfill &  $-     59.68$ &   $4.41$ &
\enddata
\end{deluxetable}

Following the first KeplerCam observation, high resolution spectroscopy
was initiated with the HIRES instrument \citep{Vogt:94} on the Keck~I
telescope, in order both to determine the stellar parameters more
precisely and to characterize the radial velocity signal due to the
companion.  With a spectrometer slit of $0\farcs86$ the resolving power
is $\lambda/\Delta\lambda \approx 55,\!000$, and the wavelength
coverage is $\sim3800-8000$~\AA{}. An iodine gas absorption cell was
used to superimpose a dense forest of $\mathrm{I}_2$ lines on the
stellar spectrum and establish a highly accurate wavelength fiducial
\citep[see][]{Marcy:92}.  In total 15 exposures were obtained between
2006 October 14 and 2007 August 25 with the iodine cell, along with one
without $\mathrm{I}_2$ for use as a template. Relative radial
velocities in the Solar System barycentric frame were derived as
described by \cite{Butler:96}, incorporating full modeling of the
spatial and temporal variations of the instrumental profile. Data and
their internal errors are listed in Table~\ref{tab:rvs}.

\section{Analysis}
\label{sec:analysis}

We determined the parameters of the star, and of the transiting planet,
from the combined photometric and spectroscopic data by the following
procedure.

First, the iodine-free template spectrum from Keck was used for an
initial determination of the atmospheric parameters of the star.
Spectral synthesis modeling was carried out using the SME software
\citep{Valenti:96}, with wavelength ranges and atomic line data as
described by \citet{Valenti:05}. We obtained initial values as
follows: effective temperature $\teff =6353 \pm 88 K$, surface gravity
$\loggstar = 3.84 \pm 0.12$, iron abundance $\mathrm{[Fe/H]}=-0.23 \pm
0.08$, and projected rotational velocity $\vsini=8.7 \pm
1.0\,\kms$. The temperature and surface gravity correspond to an F8
dwarf. The uncertainties in SME-derived parameters quoted here and in
the remainder of this discussion are twice the statistical
uncertainties; this reflects our attempt, based on prior experience,
to incorporate systematic errors.

Next, an initial modeling of the 2006 KeplerCam light curve was carried
out using the formalism based on \cite{Mandel:02}, using the quadratic
model for limb darkening {\em without} the assumption that the planet
is small.  The initial quadratic limb-darkening coefficients (namely,
$\gamma_1 = 0.1452$ and $\gamma_2 = 0.3488$) were taken from the tables
of \cite{Claret:04} by interpolation to the above-mentioned SME values. 
The period was initially fixed at the value given earlier from the
HATNet photometry, and the orbit was assumed to be circular, based on
the results from the radial velocity fit to be discussed below. The
four adjustable parameters are the planet-to-star ratio of the radii
($\rpl/\rstar$), the normalized separation ($a/\rstar$) where $a$ is the
semi-major axis of the relative orbit, the normalized impact parameter
($b \equiv a \cos i/\rstar$), and the time of the center of the transit
($T_c$).  Both for the initial light curve fit and the further fits
(see below) we used the Markov Chain Monte-Carlo (MCMC) method to find
the best fit parameters \citep[see, e.g.][for a comprehensive
description]{Ford:04}. To estimate errors we used the method of
refitting to synthetic data sets to determine their uncertainties and
correlations. The synthetic data sets consist of the analytic (fitted)
model plus Gaussian and red-noise, based on both white-noise and red-noise
estimates of the residuals. The characteristics of the red noise
component of the residuals were preserved by perturbing only the
phase of their Fourier spectrum. We found that this method of error
estimation is preferable to direct estimation using the MCMC method,
since it is not sensitive to the number of out-of-transit points used.
We also note that instead of $a/\rstar$ and $b$, we used the
parameters $\zeta/\rstar$ and $b^2$ for the fit, where $\zeta$ is an
auxiliary variable with dimensions of velocity, defined by
$\zeta\equiv2\pi a/\left(P \sqrt{1-b^2}\right)$.  These parameters have
been chosen to eliminate the correlation between $a/\rstar$ and $b^2$
\citep[see also][]{Bakos:07b}.

Next, we used the values of $\teff$ and $\mathrm{[Fe/H]}$ from the
initial SME analysis, together with the initial value of $a/\rstar$
from the Mandel-Agol fit, to estimate the stellar properties from
comparison with the Yonsei-Yale (Y$^2$) stellar evolution models by
\cite{Yi:01}.  The value of $a/\rstar$ is closely related to the
stellar density, and is thus a proxy for luminosity ($L_\star$). As
described by \cite{Sozzetti:07}, $a/\rstar$ is typically a better
constraint on $L_\star$ than the spectroscopic value of $\loggstar$,
which has a relatively subtle effect on the line profiles and whose
determination is therefore more susceptible to systematic errors.
Following \cite{Sozzetti:07} we determined the range of stellar masses
and radii that are consistent with the SME-determined values of
$\teff$, [Fe/H], and $a/\rstar$ derived from the light curve. We
obtained $\mstar = 1.19_{-0.10}^{+0.12}\,\msun$ and $\rstar =
1.45_{-0.17}^{+0.21}\,\rsun$. The resultant surface gravity,
$\loggstar = 4.19_{-0.10}^{+0.08}$, was significantly larger than the
initial SME-derived value discussed above.  We fixed $\loggstar$ at
this new, more accurate value, and repeated the SME analysis.  Fixing
surface gravity slightly affects the solution for other parameters
such as [Fe/H] and other individual element abundances as well as
$\vsini$. The correlations between these free parameters in turn
affect the final solution for effective temperature, causing it to
change by about twice its original uncertainty.  The resulting values
from this iteration are $\teff = 6570 \pm 80\,\mathrm{K}$,
$\mathrm{[Fe/H]}=-0.13 \pm 0.08$, and $\vsini = 8.2 \pm 1.0\,\kms$.

We
then performed a second iteration of the above steps using the new
values of $\teff, \log g_\star$, and $\mathrm{[Fe/H]}$, as well as
correspondingly changed limb darkening coefficents: $\gamma_1 = 0.1211$
and $\gamma_2 = 0.3646$. This iteration also incorporated both
the 2006 and 2007 KeplerCam light curves (Fig.~\ref{fig:lc}b) into a
single fit, which yielded the times of each transit center $T_{c1}$ and
$T_{c2}$, as well as the shape parameters $a/\rstar$, $\rpl/\rstar$,
and $b$, which we assumed to be the same for both of the transit events.

\notetoeditor{This is the intended place of \tabr{stellar}.}
\begin{deluxetable}{lcl}
\tablewidth{0pc}
\tablecaption{
	Stellar parameters for \hatcur{}.
	\label{tab:stellar}
}
\tablehead{
	\colhead{~~~~~~Parameter~~~~~~} &
	\colhead{Value} &
	\colhead{Source}}
\startdata
$\teff$ (K)\dotfill				&	$6570\pm80$\phn\phn	& SME\tablenotemark{a} \\
$[\mathrm{Fe/H}]$\dotfill		&	$-0.13\pm0.08$\phs  & SME \\
$\loggstar$ (cgs)\dotfill		&	$4.22\pm0.03$		& Y$^2$+LC+SME\tablenotemark{b}\\
$\vsini$ (\kms)\dotfill			&	$8.7\pm1.0$			& SME \\
$\mstar$ (\msun)\dotfill		&	$1.29 \pm 0.06$		& Y$^2$+LC+SME \\
$\rstar$ ($\rsun$)\dotfill		&	$1.46 \pm 0.06$		& Y$^2$+LC+SME \\
$\lstar$ ($\lsun$)\dotfill		&	$3.57_{-0.43}^{+0.52}$	& Y$^2$+LC+SME \\
$M_V$ (mag)\dotfill				&	$3.36\pm0.16$			& Y$^2$+LC+SME \\
Age (Gyr)\dotfill				&	$2.3_{-0.7}^{+0.5}$		& Y$^2$+LC+SME \\
Distance (pc)\dotfill			&	$260\pm20$				& Y$^2$+LC+SME
\enddata
\tablenotetext{a}{SME = Package for analysis
	of high-resolution spectra \cite{Valenti:96}.}
\tablenotetext{b}{Y$^2$+LC+SME = Yale-Yonsei isochrones \citep{Yi:01},
	light curve parameters, and SME results.}
\end{deluxetable}

Finally, using the new values of $a/\rstar$, $\teff$, and
$\mathrm{[Fe/H]}$ together with the Y$^2$ evolutionary models discussed
above, we re-determined the stellar parameters. \tabr{stellar}
summarizes these, which include a further refined determination of
$\loggstar=4.217_{-0.027}^{+0.029}$. This value is the same as the one
in the previous iteration to within uncertainties, so we stopped the
iteration at this point.

>From the value of $\vsini$ and the stellar radius $\rstar$, and
assuming an inclination of the rotational equator of approximately 90
deg (that is, similar to the inclination of the planetary orbital
plane), we estimate the stellar rotation period to be about 9 days.
We have searched the HATNet light curve (after elimination of data
during transits) for periodicities between 6 and 12 days, such as
might be produced by significant spots or other stellar activity, and
found no evidence for this.

An independent estimate of the age was obtained from the Ca$^+$ H and K
line emission strength, measured from the mean of 14 Keck spectra:
$\log R_{HK} = -4.81$, leading to an age of $2.8\,\mathrm{Gy}$
\citep[based on ][]{noyes84}, consistent with the evolutionary track
age.  The luminosity and \teff, imply an absolute visual
magnitude $M_V = 3.36 \pm 0.16$ mag.  Combined with the apparent visual
magnitude \citep[$V = 10.440 \pm 0.036$;][]{Droege:06,Hog:00}, this
yields a distance of $260 \pm 20\,\mathrm{pc}$, ignoring extinction. For
reference, the final values for the stellar parameters listed above are
given in Table~\ref{tab:stellar}.

\notetoeditor{This is the intended place of \tabr{parameters}.}
\begin{deluxetable}{lc}
\tablewidth{0pc}
\tablecaption{
	Spectroscopic and light curve solutions for \hatcur{}, and
	inferred planet parameters
	\label{tab:parameters}
}
\tablehead{
	\colhead{~~~~~~~~~~~~~~Parameter~~~~~~~~~~~~~~} &
	\colhead{Value}
}
\startdata
\noalign{\vskip -9pt}
\sidehead{Light curve parameters}
~~~$P$ (days)
\dotfill		&  $\hatcurP$  \\
~~~$T_c$ (HJD$-2,\!400,\!000$)
\dotfill		& $\hatcurMT$ \\
~~~$T_{14}$ (days)\tablenotemark{a}
\dotfill		& $\hatcurdur$ \\
~~~$T_{12} = T_{34}$ (days)\tablenotemark{a}
\dotfill		& $\hatcuringdur$ \\
~~~$a/\rstar$\dotfill					& $\hatcurar$ \\
~~~$\rpl/\rstar$\dotfill				& $\hatcurrprstar$ \\
~~~$b \equiv a \cos i/\rstar$\dotfill	& $0.602 \pm 0.030$ \\
~~~$i$ (deg)\dotfill					& $\hatcuri$ \phn \\
~~~$\zeta$ ($\rstar$/day)\dotfill	& $15.696 \pm 0.086$ \\ 
\sidehead{Spectroscopic parameters}
~~~$K$ (\ms)\dotfill               	& $\hatcurK$ \\
~~~$\gamma$ (\kms)\dotfill         	& $-22.7\pm0.5$ \\
~~~$e$\dotfill                     	& $0$ (adopted) \\
\sidehead{Planet parameters}
~~~$\mpl$ ($\mjup$)\dotfill 	& $\hatcurmlong$ \\
~~~$\rpl$ ($\rjup$)\dotfill		& $\hatcurrlong$ \\
~~~$\rho_p$ (\gcmc)\dotfill		& $\hatcurrho$ \\
~~~$a$ (AU)\dotfill                & $\hatcurarel$   \\
~~~$\gpl$ (\mss) \dotfill        & $\hatcurg$
\enddata
\tablenotetext{a}{
	\ensuremath{T_{14}}: total transit duration, time
	between first to last contact; \ensuremath{T_{12}=T_{34}}:
	ingress/egress time, time between first and second, or third and
	fourth contact.
}
\end{deluxetable}

Setting the difference between the transit centers of the two KeplerCam
light curves, $T_{c2} -T_{c1}$, to 81 orbital periods, we obtain a more
accurate value for the period: $P=\hatcurP$ days. This value, and the
reference epoch of mid-transit, $T_c \equiv T_{c1} = \hatcurT$, are
given in Table~\ref{tab:parameters}. Also given in
Table~\ref{tab:parameters} are the final fitted light curve parameters
$a/\rstar$, $\rpl/\rstar$, $b$, $i$ (orbital inclination), and the
radius of the planet, $\rpl = \hatcurrlong\,\rjup$, as determined from
$\rstar$ and $\rpl/\rstar$.
The modeled $z$-band transit light curve is shown as a solid line
superimposed on the data points in Fig.~\ref{fig:lc}b. 

\begin{figure}
\ifpdf
\plotone{img/f2.pdf}
\else
\plotone{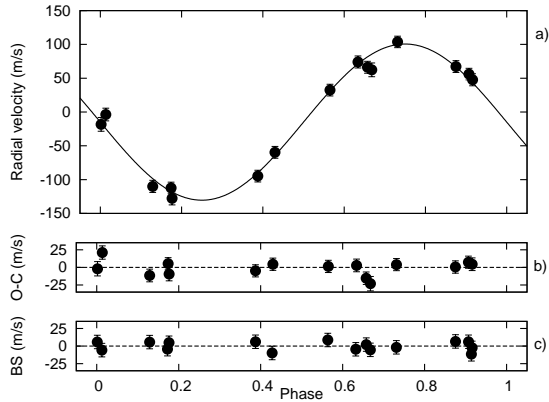}
\fi
\caption{
	(a) Radial-velocity measurements from Keck for \hatcur{}, along
	with an orbital fit, shown as a function of orbital phase. The
	center-of-mass velocity has been subtracted. (b) Phased residuals
	after subtracting off the orbital fit. The rms variation of the
	residuals is about 8.6\ms. (c) Bisector spans (BS) for the 16 Keck
	spectra plus the single template spectrum, computed as described in
	the text.  The mean value has been subtracted. Vertical scale for
	all three panels is the same.
\label{fig:rvbis}}
\end{figure}

\notetoeditor{This is the intended place of \figr{rvbis}. We would like to
typeset it as a single column figure.}

The Keck radial velocity data were initially fit with a Keplerian model
with period and epoch constrained to the value determined from the
photometry (\S\ref{sec:detection}) and no constraint on eccentricity.
We found an eccentricity of $e = 0.046 \pm 0.031$, not significantly
different from zero.  Therefore the data were re-fit with a circular
orbit, in which both the period $P$ and the mid-transit time $T_c$ were
held fixed at their values from the light curve analysis
(Table~\ref{tab:parameters}). The solution fits the data well; see
Fig.~\ref{fig:rvbis}a. No long-term trends are seen in the residuals.

In order to get a reduced chi-square value near unity for the fit, it
was necessary to add an additional random noise component with
amplitude 8.6 m s$^{-1}$ in quadrature to the internal errors.  This is
essentially identical to the 8.7 $\ms$ radial velocity ``jitter''
expected to arise from stellar surface activity, based on the
above-mentioned strength of the emission cores of the Ca$^+$ H and K
lines, $\logrhk = -4.81$ \citep{Wright:05}. The parameters of the
resulting final fit are not significantly changed by the inclusion of
this jitter, and are listed in Table~\ref{tab:parameters}.

Following \cite{Torres:07}, we explored the possibility that the
measured radial velocities are not real, but instead due to distortions
in the spectral line profiles due to contamination from a nearby
unresolved eclipsing binary.  In that case the ``bisector span'' of the
average spectral line should vary periodically with amplitude and phase
similar to the measured velocities themselves
\citep{Queloz:01,Mandushev:05}.
Instead, we detect no variation in excess of the measurement
uncertainties (see Fig.~\ref{fig:rvbis}, bottom panel). We conclude
that the velocity variations are real and that the star is orbited by a
Jovian planet.
	
Combined with the results from the light curve and radial velocity
modeling, the above stellar parameters yield a planet mass of
$\mpl=\hatcurm\,\mjup$. The surface gravity, $g_p = \hatcurg\,\ms$, was
obtained from the light curve and radial velocity fits \citep[see][]
{Southworth:07}.  The mean density and its uncertainty, $\rho_p =
\hatcurrho\,\gcmc$, follow from the absolute radius of the planet.  All
planet parameters are listed in Table~\ref{tab:parameters}.

\section{Discussion}  

\hatcurb{}, with radius $\hatcurrshort\,\rjup$, is similar in size to
five low density ``inflated'' planets tabulated by \citet{kovacs07}
(i.e.~WASP-1b, HAT-P-4b, HD~209458b, TrES-4, and HAT-P-1b).  However,
its mass of $\hatcurmshort\,\mjup$ is greater than the mass of any of
these, and hence it has a larger mean density and surface gravity. For
a planet of its mass, age of 2.3 Gy, and stellar flux $F_p$ at the
planet given by $F_p = \lstar / (4 \pi a^2)$, models of
\cite{Burrows:07} predict a radius of about 1.21 $\rjup$, assuming that
the planet has no heavy-element core. The metallicity of \hatcur{},
$\mathrm{[Fe/H]}=-0.13 \pm 0.08$, is among the smallest of known
transiting planet host stars.  If we assume that the bulk composition
of \hatcurb{} tracks the metallicity of its host star, the size of its
heavy-element core should be small, but not vanishingly so.  Thus the
predicted radius from \cite{Burrows:07} is comparable to, but perhaps
slightly higher, than one might expect for \hatcurb{}. However, the
actual radius found here, $\rpl = \hatcurrlong$, lies above the
predicted value by about $2\sigma$, so it appears to be somewhat
inflated relative to that model.

\cite{Hansen:07} proposed that hot jupiters can be placed into two
classes based on their equilibrium temperature and Safronov number
$\Theta$, where $\Theta \equiv (a/\rpl) \times (\mpl/\mstar)$ is the
ratio of the escape velocity from the surface of the planet to the
orbital velocity. When Safronov number is plotted versus equilibrium
temperature, transiting hot jupiters seem to fall into two groups, with
an absence of objects between. However, the Safronov number for
\hatcurb{} is $0.064 \pm 0.004$; along with HAT-P-5b \citep{Bakos:07b}
with Safronov number $0.059\pm0.005$, these two planets appear to fall
between the two groups in such a plot.  Hansen \& Barman also noted a
difference in the relation between planet mass and equilibrium
temperature for planets of the two classes, but HAT-P-6b and HAT-P-5b
appear to fall between the two classes in this respect as well.
It would seem that discovery and characterization of a large number of
additional transiting exoplanets may be necessary to establish
unambiguously whether there is a bi-modal distribution of hot jupiter
planets according to their Safronov number.

\acknowledgements 
Operation of the HATNet project is funded in part by NASA grant
NNG04GN74G. We acknowledge partial support also from the Kepler Mission
under NASA Cooperative Agreement NCC2-1390 (D.W.L., PI).
G.T.~acknowledges partial support from NASA under grant NNG04LG89G, and
work by G.\'A.B.~was supported by NASA through HST-HF-01170.01-A Hubble
Fellowship Grant. G.K.~and A.P.~thank the Hungarian Scientific Research
Foundation (OTKA) for support through grant K-60750. A.P.~would like to
thank the hospitality of the CfA, where this work has been partially
carried out, and acknowledge support from the Doctoral Scholarship of
E\"otv\"os University. This research has made use of Keck telescope
time granted through NASA and NOAO, of the VizieR service
\citep{Ochsenbein:00} operated at CDS, Strasbourg, France, of NASA's
Astrophysics Data System Abstract Service, and of the 2MASS Catalog.



\end{document}